\documentclass[12pt]{article}

\usepackage{amsmath}
\usepackage{amssymb}

\newtheorem{theorem}{Theorem}[section]

\newtheorem{corollary}{Corollary}[section]
\newtheorem{prop}{Proposition}[section]
\newtheorem{definition}{Definition}[section]

\numberwithin{equation}{section}

\begin{document}

\begin{center}
{\bf   A Reproducing Kernel and Toeplitz Operators 
   \\
   in the Quantum Plane}
   \vskip 1cm
   Stephen Bruce Sontz
   \\
   Centro de Investigaci\'on en Matem\'aticas, A. C.
   \\
   (CIMAT)
   \\
   Guanajuato, Mexico
\end{center}

\begin{abstract}
We define and analyze Toeplitz operators whose symbols are the elements of the
complex quantum plane, a non-commutative, infinite dimensional algebra.
In particular, the symbols do not come from an algebra of functions.
The process of forming operators from non-commuting symbols 
can be considered as a second quantization.
To do this we construct a reproducing kernel associated with the quantum plane.
We also discuss the commutation relations of creation and annihilation operators 
which are defined as Toeplitz operators.
This paper extends results of the author for the finite dimensional case.
\end{abstract}


\section{Introduction}

Based on the formalism developed in \cite{csq}, we have introduced and studied in a pair of papers
(see \cite{part1,part2}) a reproducing kernel and its associated Toeplitz operators 
which have symbols in a non-commutative algebra which is a finite dimensional
truncated version of the complex quantum plane known as a paragrassmann algebra.
We extend those results now to the case of the complex quantum plane, which
is an infinite dimensional, non-commutative algebra.
Creation and annihilation operators are defined
as certain Toeplitz operators, and their commutation relations are discussed.

This is much like a quantization scheme according 
to a common intuition of what those words should mean: ``operators instead of functions.''
However, one must modify this catch phrase to say ``operators instead of elements in a
non-commutative algebra.''
This is so because here the symbols are not elements in an algebra isomorphic to an
algebra of functions, since the latter is commutative. 
So, as we remarked in \cite{part2}, the quantization scheme discussed here is more akin to what in physics
is known as a \textit{second quantization}, where one goes
from one quantum theory to another quantum theory, rather than a \textit{first quantization},
where one goes from a classical theory to a quantum theory.

The paper is organized as follows: 
The next section introduces the basic definitions and properties.
Section~3 is about the reproducing kernel while in Section~4 we define and study
Toeplitz operators, including  the creation and annihilation operators.
Section~5 is about the commutation relations of the creation and annihilation operators.
The concluding remarks in Section~6 give some brief indications for possible future research.

\section{Definitions and such}

We study here the \textit{complex quantum plane} defined as the algebra
$$
     \mathbb{C}Q_q ( \theta, \overline{\theta} ) := 
     \mathbb{C} \{ \theta, \overline{\theta} \} / \langle \theta \overline{\theta} - q \overline{\theta} \theta \rangle
$$
where $ \mathbb{C} \{ \theta, \overline{\theta} \}$ 
is the free algebra over $\mathbb{C}$ on the two generators $\theta$ and $\overline{\theta} $ while
$\langle \theta \overline{\theta} - q \overline{\theta} \theta \rangle$ is the two sided ideal 
generated by the element $\theta \overline{\theta} - q \overline{\theta} \theta$
for some $q \in \mathbb{C} \setminus \{ 0 \}$.
This is a non-commutative algebra provided that $q \ne 1$. It has a vector space
basis
$
               AW:= \{ \theta^j \overline{\theta}{}^k \, | \, j,k \in \mathbb{N} \},
$
known as the \textit{anti-Wick basis},
and so is infinite dimensional.
In Ref.~\cite{csq} the authors call this the anti-normal ordering, which is synonymous with anti-Wick ordering.
This agrees with the definition of quantum plane in \cite{kassel} (putting the field $k = \mathbb{C}$ there)
and with the quantum $q$-plane in \cite{khal}, except for notation.
However, we will not be studying co-actions of quantum groups on this quantum space as is often done, 
but rather how its elements serve as the symbols for Toeplitz operators.

Moreover, we also define a \textit{conjugation}
(also called a \textit{$*$-operation)} in  $\mathbb{C}Q_q (\theta, \overline{\theta}) $
on the basis $AW$ by putting
\begin{equation}
\label{conjugation}
      (  \theta^j \overline{\theta}{}^k )^* :=  \theta^k \overline{\theta}{}^j 
\end{equation}
and then by extending \textit{anti-linearly} to linear combinations with coefficients in $\mathbb{C}$.
This is easily shown to be an involution, i.e., $f^{**}=f$ for all $f \in \mathbb{C}Q_q (\theta, \overline{\theta})$.
This conjugation makes $ \theta$ and $ \overline{\theta}$ into a pair of variables, each being
the conjugate of the other. 
We will see that this $*$-operation relates well with the operation of taking the adjoint
of a Toeplitz operator.
However, we are not saying (nor do we need) that this $*$-operation converts 
$\mathbb{C}Q_q (\theta, \overline{\theta}) $ into a $*$-algebra, that is $(fg)^* = g^* f^*$
for all $f,g \in \mathbb{C}Q_q (\theta, \overline{\theta}) $.
We do note without giving proof that this is a $*$-algebra if and only if 
$q \in \mathbb{R} \setminus \{ 0 \}$.

We let $w = \{ w_j \, | \, j \in \mathbb{N} \} $ be a sequence of strictly positive real numbers, that is,
$w_j > 0$. 
These will be referred to as weights.
We use these weights to define an inner product on $\mathbb{C}Q_q ( \theta, \overline{\theta} ) $
 as the sesquilinear extension
(anti-linear in the first entry, linear in the second) of
\begin{equation}
\label{ip}
    \langle \theta^a \overline{\theta}{}^b , \theta^c \overline{\theta}{}^d \rangle_w := 
    w_{a+d} \, \delta_{a+d, b+c}
    = w_{a+d} \, \delta_{a-b, c-d} \quad \mathrm{for~all~} a,b,c,d \in \mathbb{N},
\end{equation}
with $\delta$ being the Kronecker delta.
Notice that the condition $ a-b = c-d $ is necessary and sufficient for the 
inner product in (\ref{ip}) to be non-zero.
Clearly, given a pair $a,b \in \mathbb{N}$
there are infinitely many pairs $c,d \in \mathbb{N}$ such that $c - d = a - b$ and 
also satisfying $ (c,d) \ne (a,b) $.
Therefore $AW$ is not even an orthogonal basis, let alone an orthonormal basis.

We wish to note, although without giving the
relatively straightforward proof,  that there is this compatibility between the inner
product (\ref{ip}) and the conjugation (\ref{conjugation}), namely: 
$  \langle f , g \rangle_w ^* = \langle f^* , g^* \rangle_w$
for all $f, g \in\mathbb{C}Q_q ( \theta, \overline{\theta} ) $, where the $*$-operation 
on the left side is complex conjugation in $\mathbb{C}$.
We also have the identity $  \langle f , g \rangle_w ^* = \langle g , f \rangle_w$.

The definition (\ref{ip}) is partly motivated by the inner product
 introduced in \cite{csq} and studied in \cite{part1, part2}.
There one has the paragrassmann algebra defined by
$$
   PG_{l,q} (\theta, \overline{\theta}) = 
   \mathbb{C} (\theta, \overline{\theta}) / \langle \theta \overline{\theta} - q \overline{\theta} \theta, \theta^l , 
   \overline{\theta}{}^l \rangle
$$
with $l \ge 2$ an integer.
This is a quotient (as an algebra) of $\mathbb{C}Q_q ( \theta, \overline{\theta} ) $ by
the nilpotency relations $\theta^l = 0$ and $\overline{\theta}{}^l = 0$.
In that case, using the notation in \cite{part1}, 
the inner product used there satisfies
\begin{equation}
\label{ip-pg}
    \langle \theta^a \overline{\theta}{}^b , \theta^c \overline{\theta}{}^d \rangle_w 
    = \langle  \theta^{a+d}, \theta^{b+c}  \rangle 
    = w_{a+d} \, \delta_{a+d, b+c} \chi_l(a+d).
\end{equation}
Here $\chi_l$ is the characteristic function for the set of integers $\{ 0 , 1 , \dots , l-1\}$.
Its presence is due to the nilpotency relations.
Equation (\ref{ip-pg}) was not the actual definition of this inner product, although
it could have been.
Instead the definition of this inner product was given in terms of a Berezin type integral, thereby
presenting $PG_{l,q} (\theta, \overline{\theta})$ as something quite analogous to a classical
$L^2$ space.
It seems to be impossible to write (\ref{ip}) as a Berezin type integral, since now there are no
`top classes' in the theory.
However, it might be useful to express (\ref{ip}) as some sort of generalized
$L^2$ inner product.

Now another motivation for the inner product (\ref{ip}) is seen 
in the well known example of the Hilbert space
\begin{equation}
\label{theHilbertspace}
\mathcal{H}:= L^2 (\mathbb{C},  \pi^{-1} e^{-|z|^2} dx \, dy )
\end{equation}
where the monomials $z^j \overline{z}{}^k$ form a basis
(linearly independent set such that the closure of their algebraic span is the entire Hilbert space).
Then using a result that goes back at least as far to Bargmann's paper \cite{BA} in the second equality, 
for $a, b, c, d \in \mathbb{N}$ this basis satisfies
$$
        \langle z^a \overline{z}{}^b , z^c \overline{z}{}^d \rangle_{\mathcal{H}} = 
        \langle  z^{a+d}, z^{b+c}  \rangle_{\mathcal{H}} =
        (a+d)! \, \delta_{a+d, b+c},
$$
where we are using here the standard $L^2$ inner product $\langle \cdot , \cdot \rangle_{\mathcal{H}}$
in $\mathcal{H}$.
Hence we can think of $w_j$ as some sort of deformation of $j!$, the usual factorial
of $j \in \mathbb{N}$.
Notice that an immediate consequence is that $\langle z^a , \overline{z}{}^d \rangle_{\mathcal{H}} = 0 $
if either $a>0$ or $d>0$, while for $a=b=c=d=0$ we have $\langle 1, 1 \rangle_{\mathcal{H}} = 1$.
In turn this implies for $f$ holomorphic and $g$ anti-holomorphic  that
$$
     \langle f , g \rangle_{\mathcal{H}} = f(0)^* g(0).
$$
In particular such a pair of $f$ and $g$ is orthogonal if and only if
either $f(0) = 0$ or $g(0) = 0$.
This example has an interesting consequence. 
Suppose that we take the weights in the quantum plane to be $w_j  = j!$ for all $j \in \mathbb{N}$. 
Then the inner product on the quantum
plane $\mathbb{C}Q_q (\theta, \overline{\theta})$  is positive definite since in this case
$\mathbb{C}Q_q (\theta, \overline{\theta})$ is unitarily isomorphic to a dense subspace $D$ of
the Hilbert space ${\mathcal{H}}$
for any $q \in \mathbb{C} \setminus \{ 0 \}$.
In fact the map $U : \mathbb{C}Q_q ( \theta, \overline{\theta} ) \to \mathcal{H}$ 
given on the basis $AW$
by $U(\theta^i \overline{\theta}{}^j) := z^i \overline{z}{}^j$ is an isometry.
Actually $D$ is the commutative subalgebra $\mathbb{C} [z, \overline{z} ]$ of complex
polynomials in two commuting variables, and so the
unitary isomorphism 
$$
U : \mathbb{C}Q_q ( \theta, \overline{\theta} ) 
        \stackrel{\cong}{\longrightarrow} D = \mathbb{C} [z, \overline{z} ]
$$ 
 is not an algebra isomorphism for $q \ne 1$.
Also the completion of $\mathbb{C}Q_q (\theta, \overline{\theta})$ with respect to
the corresponding metric is unitarily isomorphic to the Hilbert space ${\mathcal{H}}$, again 
for any $q \in \mathbb{C} \setminus \{ 0 \}$.
And hence there are cases where the inner product defined by (\ref{ip}) is positive definite.
Motivated in part by this example
we call $\theta$ a \textit{holomorphic variable} and $\overline{\theta}$
an \textit{anti-holomorphic variable}.
(Compare also with the usage of these terms in \cite{part1} and \cite{part2}.)

However there are also cases for which the inner product (\ref{ip}) is not positive definite.
To see how this can happen, we first note some elementary calculations:
\begin{align}
\label{simple-example}
\langle 1 , 1 \rangle_w &= w_0, \nonumber
\\
\langle \theta \overline{\theta}, 1 \rangle_w &= \langle 1, \theta \overline{\theta} \rangle_w = w_1,
\\
\langle \theta \overline{\theta}, \theta \overline{\theta} \rangle_w &= w_2. \nonumber
\end{align}
As an aside, we note that $1$ is a normalized state (the `ground state') if and only if
$w_0=1$.
Let $\alpha \in \mathbb{R}$ be a real number to be specified in more detail later.
Then
\begin{equation}
\label{example}
     \langle 1 + \alpha  \theta \overline{\theta} , 1 + \alpha  \theta \overline{\theta} \rangle_w 
     = w_0 + 2 \alpha w_1 + \alpha^2 w_2,
\end{equation}
a quadratic polynomial in $\alpha$ which has distinct real roots if and only
if its discriminant is positive, that is, $ w_1^2 - w_0 w_2 > 0$.
Picking weights that satisfy this condition we see that the inner product
in (\ref{example}) will be zero for two distinct values of $\alpha  \in \mathbb{R}$ and negative for values
strictly between those two values.
(Recall that $w_2 > 0$.) 
In short, the inner product will not be positive definite in such a case.
This example also shows that $ w_1^2 - w_0 w_2 < 0$ is a necessary condition for the
inner product to be positive definite.

The remarks in the previous paragraphs show that the situation for the quantum plane
is rather different from the finite dimensional theory, 
where the inner product is never positive definite, but always non-degenerate, as shown in \cite{part1}.
We now wish to establish a necessary and sufficient condition on the weights $w_k$
so that the inner product $\langle \cdot , \cdot \rangle_w$ defined in (\ref{ip}) is non-degenerate. 
Here it is:
\begin{theorem}
\label{non-degeneracy-theorem}
The inner product (\ref{ip}) is non-degenerate on $\mathbb{C}Q_q (\theta, \overline{\theta})$
if and only if for every integer $R \ge 1$ and every $n \in \mathbb{Z}$ we have that  
$$ 
\{ \, W_{R,s} \in \mathbb{C}^R \, | \,{s \ge |n|} \, \}^\perp = 0,
$$ 
where $W_{R,s} = (w_{r+s -|n|})_{|n| \le r \le |n|+R-1}$ is a vector
in $\mathbb{C}^R$ for every $s \ge |n|$.
\end{theorem}
\textbf{Proof:}
To facilitate this argument we define a partition of the basis $AW$ so that elements in disjoint subsets
of the partition are orthogonal with respect to the inner product (\ref{ip}).
So for each integer $n  \in \mathbb{Z}$ we define
$$
     P_n := \{ \theta^a \overline{\theta}{}^b \, | \, a \ge 0, \,\, b \ge 0, \,\, a-b =n \}.
$$
Then we have $P_n \perp P_m$ for all $n,m \in \mathbb{Z}$ satisfying $n \ne m$ as well as
$$
       AW = \cup_{ n \in \mathbb{Z} } P_n,
$$
a disjoint union.
So we have an algebraic orthogonal decomposition
$$
        \mathbb{C}Q_q (\theta, \overline{\theta})  =  \oplus_{ n \in \mathbb{Z} } 
       \mathcal{P}_n,
$$
where $\mathcal{P}_n := \mathrm{span}_{\mathbb{C}} P_n$. 
(We let $\mathrm{span}_{\mathbb{C} } \, S$ denote the operation of forming the algebraic subspace over 
${\mathbb{C} } $ generated by the indicated 
set $S$.
So, we are taking here only \textit{finite} linear combinations of elements in $S$.)
It follows that the inner product (\ref{ip}) is non-degenerate if and only if it is non-degenerate
on each of the summands $\mathcal{P}_n$.

It will be convenient for us to define the  \textit{max-degree} of each basis element in $AW$ by
$$
           \mathrm{maxdeg} (\theta^a \overline{\theta}{}^b) := \max (a, b) \ge 0.
$$
Then $P_n$ contains exactly one element of max-degree $|n| +k $ for $k = 0, 1, 2 \dots$
(and no other elements).
For example, for the integers $n \le 0$ we have
$$
P_n = 
 \{ \overline{\theta}^{(-n)}  \!\! ,\, \theta \overline{\theta}^{(-n+1)} \!\! , \dots , \, 
\theta^k \overline{\theta}^{(-n+k)} \!\! , \dots \}.
$$
A similar expression holds for $n > 0$.
We denote the unique element of $P_n$ of max-degree $r$ by $\varepsilon_r$ for
each integer $r \ge |n|$.
The reader can check that for $n \ge 0$ we have $\varepsilon_r = \theta^r \overline{\theta}{}^{r-n}$,
while for $n < 0$ we have $\varepsilon_r = \theta^{r+n} \overline{\theta}{}^{r}$, where
$r \ge |n|$ in both cases.

Taking the pair of elements $\varepsilon_r, \varepsilon_s \in P_n$ for $n \in \mathbb{Z}$ 
and $r,s \ge |n|$
and then computing their inner product gives (as the reader can check) that
$$
\langle \varepsilon_r, \varepsilon_s \rangle_w = w_{r+s - |n|}.
$$
In the example (\ref{simple-example}) given earlier the two elements 
$1$ and $\theta \overline{\theta}$ lie in $\mathcal{P}_0$
and satisfy $ \mathrm{maxdeg} \,1 = 0$ and $\mathrm{maxdeg} \, \theta \overline{\theta}= 1$.
So $\varepsilon_0 = 1$ and $\varepsilon_1 = \theta \overline{\theta}$ in $\mathcal{P}_0$.

Suppose that $n \in \mathbb{Z}$ is given.
We then consider the inner product (\ref{ip}) restricted to $\mathcal{P}_n$.
Take an arbitrary element $ f \in \mathcal{P}_n $ with $f \ne 0$.
We write
$$
    f = \sum_{r \ge |n|} a_r \varepsilon_r,  
$$
where each
$a_r \in \mathbb{C}$, but only finitely many are non-zero.
But at least one of these coefficients $a_r$ is non-zero, since $f \ne 0$.
The inner product is non-degenerate on $\mathcal{P}_n$ if and only there exists $g \in \mathcal{P}_n$ 
(depending on $f$, of course) such that $\langle g ,f \rangle_w \ne 0$.
We expand $g$ as
$$
  g = \sum_{s \ge |n|} b_s \varepsilon_s 
$$
for complex coefficients $b_s$ only finitely many of which are non-zero.
Then we evaluate
\begin{equation}
\label{ip-f-g}
  \langle g ,f \rangle_w = 
     \sum_{r \ge |n|, \, s \ge |n|} a_r b_s^* \langle  \varepsilon_s, \varepsilon_r  \rangle_w
     = \sum_{s \ge |n|} b_s^*\big( \sum_{r \ge |n|} a_r w_{r+s -|n|} \big). 
\end{equation}
For example, if $w_k = 1$ (or any other constant value) for all $k \ge 0$, then taking  $f$ above 
such that $\sum_r a_r =0$ but some $a_r \ne 0$ gives us an element $f \ne 0$ satisfying
 $\langle g ,f \rangle_w = 0$ for all $g$.
So in this particular case the inner product is degenerate.

Notice that the expression in parentheses on the right in (\ref{ip-f-g}) is given to us, while the
coefficients $b_s$ are ours to choose as we please \textit{provided that} only finitely many of them
are non-zero.
So we define
\begin{equation}
\label{define-vsf}
        v_s(f) :=  \sum_{r \ge |n|} a_r w_{r+s -|n|} \in \mathbb{C}
\end{equation}
for every $s \ge |n|$.
(Recall that $n$ is a given integer so we do not include it in the notation $v_s(f)$.
The sum is well defined since only finitely many of the $a_r$'s are non-zero.)
If just one of these numbers is non-zero, say $v_{s_0}(f) \ne 0$, then we can 
put $b_s = 0 $ for all $s \ne s_0$ and $b_{s_0} = 1$. And therefore  (\ref{ip-f-g}) is non-zero.
And such a choice indeed has only finitely many (namely, one) of the $b_s's$ different from zero.
Moreover, the projection of the corresponding $g$ to $\mathcal{P}_n$, 
say $g^\prime$, satisfies $ \langle g^\prime ,f \rangle_w \ne 0$.
Therefore in this case $\{ f \}^\perp \ne \mathcal{P}_n$.
(Recall that we have restricted the inner product to $\mathcal{P}_n$.)

So if the inner product is degenerate on $\mathcal{P}_n$ (which means that $\{ h \}^\perp = \mathcal{P}_n$ 
for some $0 \ne h \in \mathcal{P}_n$), 
then there must exist some $f \ne 0$ (actually, $f=h$ works) such that $  v_s(f) = 0$ for all $s \ge |n| $.
Conversely, if there exists some $f \ne 0$ such that
$v_s(f) =0$ for all $s \ge |n|$, then for every $g$ we have $\langle g ,f \rangle_w = 0$
by (\ref{ip-f-g})
and so the inner product is degenerate on $\mathcal{P}_n$.
We now re-write the definition  (\ref{define-vsf}) for $v_s(f)$ as 
\begin{equation}
\label{rewrite}
        v_s(f) =  
        \sum_{|n| \le r \le |n|+R-1} a_r w_{r+s -|n|}\in \mathbb{C}
\end{equation}
for some integer $R \ge 1$.
Notice that the existence of $R$ is given to us implicitly as part of the information about $f$, since 
only finitely many of the $a_r$'s are non-zero.
$R$ is not unique, but that is not important for this argument. 

So we can consider $A_R(f) := (a_r^*)_{|n| \le r \le |n|+R-1}$ as a vector in $\mathbb{C}^R$.
Similarly, $W_{R,s} := (w_{r+s -|n|})_{|n| \le r \le |n|+R-1}$ 
is considered as a vector in $\mathbb{C}^R$.
Recall that $n$ is fixed since we are working in $P_n$.
However, $s \ge |n|$ is arbitrary.
We will now use the standard Hermitian inner product 
$\langle \cdot , \cdot \rangle_{\mathbb{C}^R}$ on $\mathbb{C}^R$.
Then equation (\ref{rewrite}) is the same as
$$
          v_s(f) = \langle A_R(f), W_{R,s} \rangle_{\mathbb{C}^R}.
$$
Now $\{ W_{R,s} \}_{s \ge |n|}$ is an infinite sequence of vectors in the 
finite dimensional vector space $\mathbb{C}^R$.
Since $f = \sum_r a_r \varepsilon_r$ is an arbitrary non-zero element in $\mathcal{P}_n$ with
$$
|n| \le \max (\{r \, | \, a_r \ne 0 \}) \le |n|+R-1,
$$
it follows that $A_R(f)$ is an arbitrary non-zero vector in $\mathbb{C}^R$.
Therefore the following statements are equivalent provided that $n \in \mathbb{Z}$ is given:
\begin{itemize}
\item
The inner product is degenerate on $\mathcal{P}_n$.

\item
For some $f \in \mathcal{P}_n$ with  $f \ne 0$, we have
$v_s(f) =0$ for all $s \ge |n|$.

\item
For some sequence $\{ a_r \ \, | \, r \ge |n| \}$, not identically zero but with only finitely many
terms not equal to zero, we have $v_s = 0$ for all $s \ge |n|$,
where we define $v_s := \sum_{r \ge |n|} a_r w_{r+s -|n|}$ for $s \ge |n|$.

\item
There exist some integer $R \ge 1$ and some vector $A \in \mathbb{C}^R \setminus \{ 0 \}$ such that
 for all $s \ge |n|$ we have $\langle A , W_{R,s} \rangle_{\mathbb{C}^R} = 0$. 
\item 
There exists some integer $R \ge 1$ so that
$ \{ W_{R,s} \in \mathbb{C}^R \, | \,{s \ge |n|} \}^\perp \ne 0$.
\end{itemize}
Equivalently, the inner product is non-degenerate on $\mathcal{P}_n$
if and only if 
for every integer $R \ge 1$ we have
$$
\{ W_{R,s}  \in \mathbb{C}^R \, | \,{s \ge |n|} \}^\perp = 0.
$$
We have already established that
the inner product (\ref{ip}) is non-degenerate on $\mathbb{C}Q_q (\theta, \overline{\theta})$
if and only if it is non-degenerate on $\mathcal{P}_n$ for every integer $n \in \mathbb{Z}$.
And so this finishes the proof.
$\quad \blacksquare$

\vskip 0.4cm \noindent
\textbf{Remarks:}
This result gives an algebraic necessary and sufficient condition on the weights $w_k$ for their
associated inner product to be non-degenerate.
While it looks clumsy, it should to be amenable to verification in applications.
Intuitively, the condition that an infinite sequence in a finite dimensional vector space spans the
vector space seems to be a generic condition.
And so countably many such conditions should also be generic.
Theorem \ref{non-degeneracy-theorem}
contrasts with the result for the paragrassmann algebra in \cite{part1}, where we proved that
the inner product (\ref{ip-pg}) is non-degenerate for all positive weights.

Inside the subalgebra
$$
 Pre(\theta):= \mathrm{span}_{\mathbb{C} } \{ \theta^j \, | \, j \in \mathbb{N} \} \cong \mathbb{C} [\theta]
 \subset \mathbb{C}Q_q (\theta, \overline{\theta}) 
$$
generated by all powers of the holomorphic variable $\theta$, 
we have as a particular case of the definition (\ref{ip}) that
$$
        \langle \theta^j , \theta^k \rangle_w = \delta_{j,k} w_j
$$
for all $j,k \in \mathbb{N}$. 
So the inner product restricted to the `holomorphic' subspace $Pre(\theta)$ is positive definite.
This means that $Pre(\theta)$ is a pre-Hilbert space. 
Moreover, an orthonormal basis of $Pre(\theta)$ is given by 
$$ 
     \phi_j (\theta) := \dfrac{1}{w_j^{1/2}} \, \theta^j \quad \quad \mathrm{for} \, j \in \mathbb{N}.
$$
Similar comments hold for the anti-holomorphic subalgebra $Pre (\overline{\theta})$ defined in a
completely analogous way:
$$
 Pre (\overline{\theta}):= \mathrm{span}_{\mathbb{C} } \{ \overline{\theta}^j \, | \, j \in \mathbb{N} \} 
 \cong \mathbb{C} [\overline{\theta}]
 \subset \mathbb{C}Q_q (\theta, \overline{\theta}) 
$$
We let
$$
       \mathcal{B} (\theta) = \mathcal{B} := \mathrm{comp}_{\mathbb{C} } \, Pre(\theta)
$$
denote the \textit{holomorphic space} (or the \textit{Segal-Bargmann space}) of the quantum plane.
By the operation $\mathrm{comp}_{\mathbb{C} }$ we mean the completion of the indicated pre-Hilbert space.
The set $\{ \phi_j (\theta) \, | \, j \in \mathbb{N} \}$ is also an orthonormal basis for $\mathcal{B} (\theta)$.
Unlike the finite dimensional case studied in \cite{csq, part1, part2}, the Segal-Bargmann space
$\mathcal{B} (\theta)$
here is not necessarily an algebra. 
However, it does contain the dense subspace $Pre(\theta) \cong  \mathbb{C} [\theta]$
which is an algebra, namely the algebra of polynomials in $\theta$. 
But the multiplication map for $\mathbb{C} [\theta]$ is not necessarily continuous 
in the topology induced by the norm associated to the inner product (\ref{ip})
and, if that is the case, then 
it is not extendible by continuity to $\mathcal{B} (\theta) $.

Analogously, we define the \textit{anti-holomorphic space} (or the \textit{anti-Segal-Bargmann space})
of the quantum plane by
$$
       \mathcal{B} (\overline{\theta}) := \mathrm{comp}_{\mathbb{C} } \, Pre (\overline{\theta}).
$$
These two spaces  $\mathcal{B} (\theta) $ and $\mathcal{B} (\overline{\theta})$ should be understood as `almost' disjoint.
Their `intersection' is the one dimensional subpace spanned by $1 = \theta^0 = \overline{\theta}{}^0$.

\section{Reproducing kernel}

As a first step towards the definition of Toeplitz operators, we shall find a reproducing kernel for
the Segal-Bargmann space.
First off, we will need to define a functional calculus for the Segal-Bargmann space.
As is well-known, there always is a functional calculus for polynomials $f \in \mathbb{C}[x]$
associated to any element in any associative algebra. 
Here we write 
$$
     f = \sum_{j=0}^m a_j x^j  \in \mathbb{C}[x]
$$
with coefficients $a_j \in \mathbb{C}$ and then use the standard definition
$$
      f(\theta) := \sum_{j=0}^m a_j \theta^j.
$$
But there are some elements in $\mathcal{B} (\theta)$ that are not so representable,
since they are infinite sums of elements in the orthogonal basis $\{ \theta^j \}$.
However, any element $u \in \mathcal{B} (\theta)$ can be expanded as an infinite
sum with respect to the orthonormal basis $\{ \phi_j (\theta) \}$ giving
$$
           u = \sum_{j=0}^\infty a_j \phi_j (\theta) = \sum_{j=0}^\infty a_j w_j^{-1/2} \theta^j
$$
with $a_j \in \mathbb{C}$ and $\sum_j |a_j|^2 < \infty$.
Equivalently, for all $u \in \mathcal{B} (\theta)$ we have
$$
          u = \sum_{j=0}^\infty f_j \theta^j
$$
with $f_j \in \mathbb{C}$ and $\sum_j | f_j |^2 w_j < \infty$.
So associated to any sequence of positive real numbers 
$w = \{ w_j \, | \, j \ge 0  \}$ we define a weighted little $l^2$ space:
$$
            l^2 (w) := \{ \, f =\{ f_j \, | \, j \in \mathbb{N} \} \, \, \big| \, \sum_j | f_j |^2 w_j < \infty \, \}.
$$
Then the full \textit{functional calculus of $\theta$} is the linear mapping
$$
        \Phi : l^2 (w) \to   \mathcal{B} (\theta) 
$$
defined by $\Phi ( f ) = \Phi ( \{ f_j \} ) := \sum_{j=0}^\infty f_j \theta^j$.
So $\Phi$ is a unitary isomorphism of Hilbert spaces.
We also use the more suggestive notation $f (\theta ) := \Phi (f )$
for all $f \in l^2 (w)$.

Now the reproducing kernel $K (\theta, \eta)$ is supposed to satisfy the \textit{reproducing kernel formula}, namely
\begin{equation}
\label{repro-formula}
      f (\theta) = \langle K (\theta, \eta) , f(\eta) \rangle_w \in \mathcal{B} (\theta)
\end{equation}
for all $f \in  l^2 (w)$ and where $\eta \in \mathcal{B} (\eta)$ is another `independent
copy' of a holomorphic variable. 
The intuitive idea behind the inner product in (\ref{repro-formula})
is that it should only take $\eta$ into consideration 
while letting $\theta$ have a free ride as a `passenger'.
The usual structure of reproducing kernel functions in spaces of holomorphic functions
suggests that we should have 
$$
     K (\theta, \eta) \in \mathcal{B} (\overline{\theta}) \otimes \mathcal{B} (\eta),
$$
the standard tensor product of Hilbert spaces.
This expresses the intuition that $K (\theta, \eta) $ should be anti-holomorphic
in $\theta$ and holomorphic in $\eta$.
So we want to define an inner product $\langle L (\theta, \eta) , f(\eta) \rangle_w$
for all $L  (\theta, \eta) \in \mathcal{B} (\overline{\theta}) \otimes \mathcal{B} (\eta)$ and
all $f \in  l^2 (w)$.
Actually, we will start off this discussion by suppressing the Hilbert space structures and
simply considering 
$ f(\eta) =\sum_k  f_k \eta^k $, a formal infinite sum, and
$$
            L  (\theta, \eta) = \sum_{ij} \lambda_{ij} \, \overline{\theta}{}^i \otimes \eta^j,
$$
another formal infinite sum (that is, no convergence requirements).
We now make the following formal calculation in order to motivate a definition:
\begin{align}
\label{define-extended-ip}
            \langle L (\theta, \eta) , f(\eta) \rangle_w &= 
            \sum_{ijk} \lambda_{ij}^* f_k \langle  \overline{\theta}{}^i \otimes \eta^j, \eta^k    \rangle_w
            \nonumber
            \\
            &= 
            \sum_{ijk} \lambda_{ij}^* f_k \langle \eta^j, \eta^k    \rangle_w  \theta^i 
            \nonumber
            \\
             &= 
            \sum_{ijk} \lambda_{ij}^* f_k \delta_{j,k} w_j \theta^i
            \nonumber
            \\
             &= 
            \sum_{i} \big( \sum_j \lambda_{ij}^* f_j w_j \big) \theta^i.
\end{align}
The inner sum in (\ref{define-extended-ip}) over $j \ge 0$ is an infinite sum of complex numbers 
for every $i \ge 0$ and so will not be considered as a formal infinite sum.
But to consider it as an absolutely convergent series, say, 
we will have to impose conditions on the coefficients $\lambda_{ij}$
and $f_k$ of the above formal expressions.
(The weights $w_j$ are considered as given.) 
After all the inner sums in (\ref{define-extended-ip})
have been well defined we are left with a formal expression,
namely a formal power series in the variable $\theta$.
This can be used as such.
Or, if one prefers, some more conditions can be imposed so that this series
converges in some topological vector space, which could be $\mathcal{B}(\theta)$ with
one of its many topological structures (norm topology, weak toplogy, etc.).

For example, we can use H\"older's inequality to get the estimate
\begin{equation}
\label{holder}
       \sum_j | \lambda_{ij}^* f_j w_j | \le \big( \sum_j   | \lambda_{ij}|^p  w_j\big)^{1/p}  
        \big( \sum_j |f_j |^{p^\prime} w_j \big)^{1/p^\prime} 
\end{equation}
for any $1 < p < \infty$, where $p^\prime$ is the usual dual index of $p$.
Consequently, if there exists some $1 < p < \infty$ such that the first sum on the
right side of (\ref{holder}) is finite for every $i \ge 0$ and such that the second sum is finite,
then we have that the formula (\ref{define-extended-ip}) defines the inner product 
$\langle L (\theta, \eta) , f(\eta) \rangle_w$ as a formal power series in $\theta$. 

We next consider the canonical orthogonal basis of $ l^2 (w)$ given by 
$$\varepsilon_j = ( 0, \dots, 0 , 1, 0, \dots )$$ (all zeros with one single occurrence 
of $1$ in entry $j \in \mathbb{N}$).
Then we have 
$$
   \varepsilon_j (\theta) = \Phi (\varepsilon_j) = \theta^j.
$$
So a necessary 
condition for (\ref{repro-formula}) to hold is that
\begin{equation}
\label{special-case}
         \theta^j = \langle K (\theta, \eta) , \eta^j \rangle_w
\end{equation}
for all $ j \in \mathbb{N}$.
We look for a solution
$K (\theta, \eta) = \sum_{kl} a_{kl} \overline{\theta}{}^k \otimes \eta^l$,
a formal series,
for unknown coefficients $a_{kl} \in \mathbb{C}$.
So we use our formal definition (\ref{define-extended-ip}) to get 
\begin{equation*}
\langle K (\theta, \eta) , \eta^j \rangle_w =
\sum_{k} a_{kj}^* \, w_j \theta^k,
\end{equation*}
a formal power series in $\theta$.
So (\ref{special-case}) holds if and only if 
\begin{equation}
\label{explicit-special-case}
       \theta^j = \sum_{k} a_{kj}^* w_j \, \theta^k  
\end{equation}
for all $ j \in \mathbb{N}$.
Of course, the left side of (\ref{explicit-special-case}) is a \textit{finite} series.
Clearly, (\ref{explicit-special-case}) is satisfied if and only if $a_{jk} = \delta_{j,k} / w_j$.

Putting this into the formula for the reproducing kernel gives us
\begin{align}
\label{formula-for-K}
         K (\theta, \eta) &= \sum_{kl} a_{kl} \overline{\theta}{}^k \otimes \eta^l 
         = \sum_{kl} \dfrac{\delta_{k,l} }{ w_k } \overline{\theta}{}^k \otimes \eta^l 
         = \sum_k \dfrac{1}{w_k} \overline{\theta}{}^k \otimes \eta^k  \nonumber
         \\
         &= \sum_k \phi_k(\overline{\theta}) \otimes \phi_k(\eta).
\end{align}
But this series is not convergent in the norm topology of the Hilbert space
$ \mathcal{B} (\overline{\theta}) \otimes \mathcal{B} (\eta)$, since the terms satisfy
$$
         || \phi_k (\overline{\theta} ) \otimes  \phi_k(\eta) ||
        = 1.
$$
However, there is another topology on $ \mathcal{B} (\overline{\theta}) \otimes \mathcal{B} (\eta)$
for which this series is convergent.
This other topology corresponds to the strong operator topology (see \cite{reed-simon})
in the space
$\mathcal{L} (\mathcal{B}(\eta), \mathcal{B}(\theta))$ of bounded linear operators
mapping $\mathcal{B}(\eta)$ to $\mathcal{B}(\theta)$.
Without going into a lot of technical details, let us simply note that the formula 
(\ref{formula-for-K}) induces a unitary isomorphism $S: \mathcal{B}(\eta) \to \mathcal{B}(\theta)$
given in Dirac notation by
$$
         S = \sum_k | \phi_k (\theta) \rangle \langle \phi_k (\eta) |
$$
which is an infinite sum of rank one operators, each of which has operator norm $1$,
and so is not convergent in the operator norm topology.

Nonetheless this infinite series of operators is convergent in the strong operator topology.
It satisfies $S : \phi_k (\eta) \mapsto  \phi_k (\theta)$ for the basis elements
and so $S : f (\eta) \mapsto  f (\theta)$ 
for $f \in l^2(w)$.
This is quite tautological, since this is exactly what the mapping induced
by the reproducing kernel, as given by the right side of equation (\ref{repro-formula}),
 is supposed to do! 
 And so it does.
Intuitively, the expression in (\ref{formula-for-K}) expresses in this context the formula
for the kernel of the Dirac delta as a `smooth' object.

This section may seem like a lot of work to arrive at a result that appears to lack substance.
However, the formula (\ref{formula-for-K}) will be used in the next section to define
Toeplitz operators in a rather natural way.
And these Toeplitz operators have some substantial, non-trivial properties. 
There may be other ways, still to be discovered, for defining these Toeplitz operators.
But for the time being we seem to have found a reasonable  approach.

Also, it is worth mentioning that the reproducing kernel $K$ in (\ref{formula-for-K})
is not a function of two variables in the usual sense of those words. 
If it were, then $f(\theta)$ would be the `value' of $f$ at the `point' $\theta$.
But $f(\theta)$ is an element in $\mathcal{B}(\theta)$ for all $f \in l^2(w)$.
And $\theta$ itself is an element in the very same space $\mathcal{B}(\theta)$.
So the sort of reproducing kernel as given in (\ref{formula-for-K}) is not included 
in the classical theory of reproducing kernel \textit{functions} such as found in 
\cite{aron} and \cite{saitoh}.
For example, the usual point-wise estimate, which follows immediately from the 
Cauchy-Schwarz inequality in the classical case, seems to have no good analogue here.
Anyway, the Cauchy-Schwarz inequality does not apply to the general reproducing kernel formula
in (\ref{repro-formula}) nor to its special case (\ref{special-case}).

But there are some properties of the reproducing kernel (\ref{formula-for-K}) 
that are analogous to standard properties of reproducing kernel functions. 
(See \cite{aron} and \cite{saitoh}.)
The correct interpretation of the following properties entails defining with some care 
notations which superficially appear obvious. 
We will not go into that analysis, but refer the reader to \cite{part1} where a similar
analysis was made.
We now present these properties:
\begin{enumerate}

\item Positive definite: $\sum_{n,m=1}^N \lambda_ n^* \lambda_m K (\theta_n, \theta_m) \ge 0 \,$
for $\, \lambda_1, \dots , \lambda_N \in \mathbb{C}$.

\item Complex symmetry: $K(\theta, \eta)^* = K(\eta, \theta)$.

\item Self reproducing: $K(\theta, \eta) = \langle K(\eta, \cdot), K(\theta, \cdot) \rangle_w$.

\item Positivity on the diagonal: $K(\theta, \theta) =  \sum_k | \phi_k (\theta) \rangle \langle \phi_k (\theta) |
= I_{\mathcal{B} (\theta)} \ge 0$.

\end{enumerate}

Also, there is the question of constructing a space with a given $K(\theta, \eta)$ (satisfying  properties 1 and 2)
 as its reproducing kernel.
 While this is a well known result in the theory of reproducing kernel \textit{functions}, it appears that
 the analogous construction can not be made here since we are not dealing with functions.

\section{Toeplitz Operators}

Much of the above material about the reproducing kernel appears to be somewhat
tautological in nature, though with a lot of technical details since here we are dealing 
with infinite dimensional spaces rather
than the finite dimensional theory in \cite{part1}.
But the real point of the reproducing kernel for us is that it can be extended in a `natural'
manner to the quantum plane and as such becomes one of the principle ingredients
in defining a non-trivial theory of Toeplitz operators with symbols in the complex quantum plane,
a non-commutative algebra for $q \ne 1$.
As noted earlier in \cite{part2}, passing from a symbol in a non-commutative algebra 
to its Toeplitz operator is an example of second quantization, since it is the
quantization of a theory that is itself a non-commutative (i.e., quantum) theory
to begin with.
Nonetheless, the initial theory is still often referred to as the classical theory.

To start off this discussion we define the inner product of any finite sum
or any infinite (formal) sum of the form
$ M(\theta, \eta) = \sum_{jk} m_{jk} \, \overline{\theta}^j \otimes \eta^k$
with coefficients $m_{jk} \in \mathbb{C}$ for $j,k \ge 0$
and a basis element  $\eta^a \overline{\eta}{}^b \in \mathbb{C}Q_{q}(\eta, \overline{\eta})$
in $AW$ by
\begin{align}
\label{prelim-ip}
 \langle M(\theta, \eta) \, , \,  \eta^a \overline{\eta}{}^b \rangle_w &:=
 \sum_{j} \left( \sum_k m_{jk}^* \langle \eta^k , \eta^a \overline{\eta}{}^b \rangle_w  \right) \theta^j
 \nonumber
 \\
 &= 
 \sum_{j} \left( \sum_k m_{jk}^* \langle \eta^{k+b} , \eta^a \rangle_w  \right) \theta^j
 \nonumber
 \\
 &= 
 \sum_{j} \left( \sum_k m_{jk}^* \delta_{k+b,a} w_a  \right) \theta^j
 \nonumber
 \\
 &= 
  w_a \sum_{j} m^*_{j, a-b}  \, \theta^j
\end{align}
provided that 
the sum on $j$ converges in $\mathcal{B}(\theta)$, which is equivalent to 
$$\sum_j w_j| m_{j, a-b} |^2 < \infty.$$
Or we could simply take (\ref{prelim-ip}) to be a formal series.
Here we have introduced the convention that $m_{jk} = 0$ if $k < 0$.
Then for any given arbitrary element
$F = \sum_{ab} c_{ab} \eta^a \overline{\eta}{}^b \in  \mathbb{C}Q_{q}(\eta, \overline{\eta})$
(which is always a finite sum)
such that for each pair $(a,b)$ satisfying $c_{ab} \ne 0$ we have convergence in (\ref{prelim-ip}),
we define 
$$
         \langle M(\theta, \eta) \, , F \rangle_w := 
         \sum_{ab} c_{ab} \langle M(\theta, \eta) \, , \,  \eta^a \overline{\eta}{}^b \rangle_w,
$$
which is also a finite sum.
Notice that this inner product in general takes values in $\mathcal{B}(\theta)$ provided that
we impose the convergence conditions, though
in some specific cases the inner product could lie in some subspace of $\mathcal{B}(\theta)$.
 
Next we define the operator associated with the reproducing
kernel $K$.
This is the extension of the reproducing kernel to the quantum plane that we mentioned earlier.
\begin{definition}
The operator associated to the reproducing kernel of $Pre(\theta)$,
$P_K : \mathbb{C}Q_{q}(\theta, \overline{\theta}) \to \mathbb{C}Q_{q} (\theta, \overline{\theta})$,
is defined by
\begin{equation}
\label{define-PK}
P_K F (\theta) := \langle K(\theta, \eta) , F(\eta, \overline{\eta}) \rangle_w
\end{equation}
for all $ F(\theta, \overline{\theta}) \in \mathbb{C}Q_{q}(\theta, \overline{\theta})$.
\end{definition}
This definition comes down to a special case of the discussion in the previous paragraph.
So we must show that the inner product in (\ref{define-PK}) is well defined.
Also $P_K$ is actually a symmetric projection as we prove next.

\begin{theorem}
$P_K$ is well defined and is a projection, that is, $P_K^2 = P_K$.
Also, $P_K$ is symmetric with respect to the inner product $\langle \cdot , \cdot \rangle_w$.
\end{theorem}
\textbf{Remark:} Since this inner product is not necessarily non-degenerate, we do
not always have that the adjoint of $P_K$ exists.
Nonetheless, it makes sense to speak of the symmetry of $P_K$.
And in those cases when the inner product is non-degenerate, we do
have $P_K^* = P_K$.
\vskip 0.4cm \noindent
\textbf{Proof:}
We write
$F_{ab} (\theta, \overline{\theta}) := \theta^a \overline{\theta}{}^b$
for the elements in the basis $AW$.
We extend the notation established above by setting $\theta^n =0$ and $w_n = 1$
for all integers $n < 0$.
As we noted earlier, this basis $AW$ is not orthogonal.

Acting with $P_K$ on the basis elements $F_{ab}$ in $AW$ we obtain
\begin{align}
\label{act-on-AW}
&(P_K F_{ab}) (\theta) =  \langle K(\theta, \eta) , F_{ab} (\eta, \overline{\eta}) \rangle_w
=  \langle K(\theta, \eta) , \eta^a \overline{\eta}{}^b \rangle_w \nonumber
\\
&= \sum_j \dfrac{1}{w_j} \langle  \eta^j , \eta^a \overline{\eta}{}^b \rangle_w \, \theta^j
= \sum_j \dfrac{1}{w_j} \delta_{j+b,a} w_a \, \theta^j 
= \dfrac{w_a}{w_{a-b}} \theta^{a-b}.
\end{align}
This result corresponds in this case to the convergence in (\ref{prelim-ip}) for
all $a,b$.
In this particular case, the infinite series collapses to at most one non-zero term, and so
we have convergence not only in $\mathcal{B} (\theta)$
but even to an element in its subspace $Pre(\theta)$.
So by extending linearly to finite sums we see that the definition (\ref{define-PK})
makes sense.
Moreover, (\ref{act-on-AW}) shows that $Ran \, P_K \subset Pre(\theta)$.
In particular by putting $b=0$ in (\ref{act-on-AW})
we find that $(P_K F_{a,0}) (\theta) = F_{a,0}(\theta) $,
which says $P_K : \theta^a \mapsto \theta^a$, that is, $P_K$ acts as the identity on $Pre(\theta)$.
So, $P_K^2 = P_K$ and $Ran \, P_K = Pre(\theta)$ follow immediately.

For the symmetry of $ P_K $ we calculate various matrix elements for $P_K$ 
with respect to the elements in the basis $AW$. 
First for $P_K$ acting on the right entry we obtain:
\begin{align}
\label{acting-on-right}
\langle F_{ab}, P_K F_{cd} \rangle_w &=
\langle \theta^a \overline{\theta}{}^b, \dfrac{w_c}{w_{c-d}} \theta^{c-d}
\rangle_w 
= \dfrac{w_c}{w_{c-d}}    \delta_{a, b+c-d}  w_a  H(c-d) \nonumber
\\
&= \dfrac{w_a w_c}{w_{c-d}}  \delta_{a-b,c-d} H(c-d) ,
\end{align}
where $H$ is the (discrete) Heaviside function $H: \mathbb{Z} \to \{ 0, 1 \}$
defined by $H(n) := 1$ for $n \ge 0$ and $H(n) := 0$ for $n < 0$.

Next we calculate the matrix elements for $P_K$ acting on the left entry:
\begin{align}
\label{acting-on-left}
 \langle P_K F_{ab},  F_{cd} \rangle_w &=
\langle \dfrac{w_a}{w_{a-b}}  \theta^{a-b}, \theta^c \overline{\theta}{}^d \rangle_w 
= \dfrac{w_a}{w_{a-b}}  H(a-b) \delta_{a-b+d,c}  w_c \nonumber
\\
&= \dfrac{w_a w_c}{w_{a-b}} H(a-b) \delta_{a-b,c-d}.
\end{align}
Since the matrix entries (\ref{acting-on-right}) and (\ref{acting-on-left}) 
with respect to the elements in the vector space
basis $AW$ of $\mathbb{C}Q_{q}(\theta, \overline{\theta})$ are equal, 
we can pass to finite linear combinations to get
$$
\langle F, P_K G \rangle_w = \langle P_K F , G \rangle_w
$$
for all $F,G \in \mathbb{C}Q_{q}(\theta, \overline{\theta}) $,
which is the desired symmetry of $P_K$.
$\quad \blacksquare$

\vskip 0.4cm
Because of the previous proof we can think of $P_K$ as a mapping
$$
P_K : \mathbb{C}Q_{q}(\theta, \overline{\theta}) \to Pre(\theta) \subset  \mathcal{B} (\theta).
$$
For any $g \in \mathbb{C}Q_{q}(\theta, \overline{\theta})$ we define the linear map 
$M_g : Pre(\theta) \to \mathbb{C}Q_{q}(\theta, \overline{\theta})$
to be multiplication by $g$ on the right, that is
$$
        M_g \phi := \phi g
$$
for all $\phi \in Pre(\theta)$.
It is straightforward to show that $Ran \, M_g \subset \mathbb{C}Q_{q}(\theta, \overline{\theta})$.
\begin{definition}
We define the {\em Toeplitz operator} associated to the {\em symbol}
$g \in \mathbb{C}Q_{q}(\theta, \overline{\theta})$ to be
$$
T_g = P_K M_g,
$$
that is, right multiplication by $g$ followed by the projection associated to
the reproducing kernel $K$.
We also write
$$
            T_g : Pre(\theta) \to  \mathcal{B} (\theta) 
$$
with the domain of $T_g$ defined by $Dom (T_g) := Pre(\theta) \subset \mathcal{B} (\theta)$
to indicate that $T_g$ is a densely defined linear operator acting in (but not on)
the Segal-Bargmann space $\mathcal{B} (\theta)$.

An equivalent way to write this definition is
$$
T_g f (\theta)  = \langle K(\theta, \eta) \, , \,  f(\eta) \, g(\eta, \overline{\eta}) \rangle_w ,
$$
where $f \in Pre(\theta)$.

\end{definition}

Actually, we have that $Ran \, T_g \subset Pre(\theta)$, but we prefer to consider the
codomain to be the larger space $\mathcal{B} (\theta)$ in order to be able to apply
the theory of densely defined linear operators acting in a Hilbert space.
For example, see \cite{reed-simon}.
The definition of $T_g$ can be expressed as this composition:
\begin{equation*}
Dom (T_g) =
Pre(\theta) \stackrel{M_g}{\longrightarrow} \mathbb{C}Q_q(\theta, \overline{\theta})
\stackrel{P_K}{\longrightarrow} Pre(\theta) \subset \mathcal{B} (\theta)
\end{equation*}

One of the first considerations here is to find necessary and sufficient conditions
on $g$ in order that $T_g$ is bounded and so has a unique bounded extension
to $ \mathcal{B} (\theta)$.
And when $T_g$ is bounded, one would like some information, at best a formula but
at least an estimate,
about the operator norm of $T_g$.
While bounded operators are important, we will also be interested in certain operators that are not bounded.

We have used the common way of defining Toeplitz operators:
multiply by a symbol and then project back into the Hilbert space.
However, we are making choices here that are somewhat arbitrary.
For example, we could have used left multiplication instead of right
multiplication.
Also the choice of the Segal-Bargmann space is arbitrary too.
We could just as well have chosen the anti-Segal-Bargmann space, which
also has a reproducing kernel.
And having chosen instead that space, we would again have two
possible choices for the multiplication operator: left and right.
In all, there are four different choices for the definition of Toeplitz operators,
and we simply have opted for one of these. 
The other three choices lead to very similar theories and will not be discussed further.

Next we define the {\em Toeplitz mapping} $T: g \mapsto T_g$ giving us a linear function
$$
T : \mathbb{C}Q_{q}(\theta, \overline{\theta}) \to \mathcal{L} ( \mathcal{B}(\theta) : Pre (\theta) ),
$$
where $\mathcal{L} ( \mathcal{B}(\theta) : Pre (\theta)) $ is the
complex vector space of all linear densely defined
operators $S$ acting in the Hilbert space $\mathcal{B}(\theta)$ with $Dom \, S  = Pre (\theta)$
and leaving $Pre (\theta)$ invariant, that is
$S(Pre (\theta) ) \subset Pre (\theta)$.
Because of this last condition $\mathcal{L} ( \mathcal{B}(\theta) : Pre (\theta)) $ is closed under
composition and so is an algebra.
We also call $T$ the {\em Toeplitz quantization}.

One verifies that $T_1 = I_{Pre (\theta)}$, the identity,
as an immediate consequence of the fact that $K$ is the reproducing kernel of $Pre (\theta)$.
However, even though $T$ is a map from one algebra to another algebra, it is not an algebra
morphism. 
The product on the domain space is determined by $q \in \mathbb{C} \setminus \{ 0 \}$, while the
operator $T_g$ is defined using the inner product which depends on the weights $w_k$.
Even when the weights are functions of $q$ (and so are not independent quantities) it is not
expected that $T$ preserves products, given what happens with Toeplitz operators in other contexts.
Here is a result which shows what is happening in a `nice' case.
\begin{theorem}
\label{nice-theorem}
Suppose that we have symbols $g_1$ and $g_2$, but with $g_2 \in Pre(\theta)$, that is
$g_2$ `depends' only on $\theta$.
Then $T_{g_1} T_{g_2} = T_{g_2 g_1}$. 
\end{theorem}
\textbf{Proof:}
The point is  since $g_2 \in Pre(\theta)$ we have that $T_{g_2} = P_K M_{g_2} = M_{g_2}$,
because multiplication by $g_2$ leaves $Pre(\theta)$ invariant.
So we calculate
\begin{align*}
T_{g_1} T_{g_2} &=   P_K M_{g_1}  P_K M_{g_2} =  P_K M_{g_1}  M_{g_2}  =  P_K M_{g_2 g_1}
= T_{g_2 g_1},
\end{align*}
where the second to last equality is left to the reader to check.
$\quad \blacksquare$
\vskip 0.4cm \noindent
\textbf{Remark:}
In the standard theories of Toeplitz operators, the symbols are functions and so commute. 
So essentially the same argument in such cases (with the corresponding hypothesis!) gives 
 $T_{g_1} T_{g_2} = T_{g_1 g_2}$.
 The fact that the map $T$ in this context reverses the order of multiplication in this special
 case is not important as such. 
 The equation $P_K M_{g} = M_{g}$
is not true for all symbols $g$ and this is what is behind the fact that $T$ does not respect multiplication.
In fact, Theorem \ref{nice-theorem} implies that 
$T_{\overline{\theta}} T_{\theta} = T_{\theta {\overline{\theta}} }$.
In the next calculation we actually will use something ever so slightly stronger, namely 
$T_{\overline{\theta}} T_{\theta} = T_{\theta {\overline{\theta}} }\ne 0$, but this will become clear later on.
So for $q \ne 1$ we have
$$
       T_{\overline{\theta} \theta } = T_{q^{-1} \theta {\overline{\theta}} }
   = q^{-1}  T_{\theta {\overline{\theta}} }  = 
   q^{-1}  T_{\overline{\theta}} T_{\theta} \ne T_{\overline{\theta}} T_{\theta}.
$$
Later on we will also calculate $T_{\theta} T_{\overline{\theta}}$ and see that this is yet another
operator also not equal, in general, to $T_{\overline{\theta} \theta }$.

\begin{theorem}
\label{T-is-monomorphism}
The linear map
$
T : \mathbb{C}Q_{q}(\theta, \overline{\theta}) \to \mathcal{L} ( \mathcal{B}(\theta) : Pre (\theta) )
$
is a vector space monomorphism  if and only if the inner product (\ref{ip}) is non-degenerate.
\end{theorem}
\textbf{Proof}:
We are looking for a necessary and sufficient for $\ker \, T =0$.
So we take $g \in \ker \, T$, which means that  $T_g = 0$.
In particular, this is equivalent to $T_g f_d = 0$ for all $d \ge 0$, where $f_d = \theta^d$,
an orthogonal basis of $Pre(\theta)= Dom (T_g)$.
We calculate
\begin{align*}
T_g f_d (\theta)  &= \langle K(\theta, \eta) \, , \,  f_d(\eta) \, g(\eta, \overline{\eta}) \rangle_w
\\
&=   \sum_c  \dfrac{1}{w_c} \langle \overline{\theta}{}^c \otimes \eta^c, \eta^d g(\eta, \overline{\eta})   \rangle_w
 \\
&=  \sum_c  \dfrac{1}{w_c} \langle \eta^c  \overline{\eta}{}^d , g(\eta, \overline{\eta})  \rangle_w \, \theta^c.
\end{align*}
So, $T_g f_d (\theta) = 0$
for all $d \ge 0$ if and only if
$  \langle \eta^c  \overline{\eta}{}^d \, , g(\eta, \overline{\eta})  \rangle_w =0$
for all $c, d \ge 0$ if and only if $g(\eta, \overline{\eta}) $ is orthogonal to 
$ \mathbb{C}Q_{q}(\theta, \overline{\theta}) $.
So $\ker \, T = \big(    \mathbb{C}Q_{q}(\theta, \overline{\theta})    \big)^\perp$ and the
result follows. $\quad \blacksquare$

\vskip 0.4cm
\textbf{Remarks:}
One way to interpret this theorem is that it tells us when the symbol of a Toeplitz operator
is uniquely determined by the operator.
In the finite dimensional theory presented in \cite{part1} the inner product is always non-degenerate
and the corresponding result proved there is that the Toeplitz quantization is always a monomorphism.
Moreover in the context of \cite{part1} the domain and codomain vector space of the 
Toeplitz quantization have the same \textit{finite} dimension;
therefore that Toeplitz quantization is automatically a vector space (but not algebra) isomorphism.
Here one expects the situation to be more complicated due to the fact that the domain and
codomain of $T$ have infinite dimension.
To be more precise one expects that $T$ is not surjective, that is, there exist operators which
are not Toeplitz.
Moreover, in the current context Toeplitz operators are not necessarily bounded as we shall
see momentarily.

\vskip 0.4cm
We calculate next the Toeplitz operators for the basis elements
$\theta^i \overline{\theta}{}^j$ of the symbol space $PG_{l,q}(\theta, \overline{\theta}) $.
\begin{theorem}
The action of the Toeplitz operator $T_{\theta^i \overline{\theta}{}^j}$ on the orthonormal
basis elements  $\phi_a(\theta) = w_a^{-1/2} \theta^a \in Pre(\theta)$ with $a \ge 0$
is given by
\begin{equation}
   \label{useful}
    (T_{\theta^i \overline{\theta}{}^j } \phi_a) (\theta) = \dfrac{w_{i+a}}{ (w_a \, w_{i+a-j} )^{1/2} } \, \phi_{i+a-j}(\theta) .
\end{equation}
\end{theorem}
\textbf{Proof:}
We evaluate as follows:
\begin{align*}
  (T_{\theta^i \overline{\theta}{}^j } \phi_a)    (\theta) &=
    \langle K(\theta, \eta) \, , \, \phi_a(\eta) \, \eta^i \overline{\eta}{}^j    \rangle_w
    \\
      &= \langle \sum_{k} \phi_k ( \overline{\theta} ) \otimes \phi_k (\eta) \, , \, 
   w_a^{-1/2} \eta^a \eta^i \overline{\eta}{}^j  \rangle_w 
   \\ 
   &=
   w_a^{-1/2} \sum_{k}  w_k^{-1/2} \langle \eta^{j+k} \, , \, \eta^{i+a}  \,  \rangle_w \, \phi_k (\theta)
    \\
      &= w_a^{-1/2} \sum_{k}  w_k^{-1/2} \delta_{j+k, i+a} w_{j+k} \, \phi_k (\theta) 
    \\
   &=
     \dfrac{w_{i+a}}{ (w_a \, w_{i+a-j} )^{1/2} } \, \phi_{i+a-j}(\theta) .
\end{align*}
Recall that $\theta^n =0$ and $w_n = 1$ for $n < 0$.
So we also have put $\phi_n (\theta) = 0$ for $ n < 0$
in the above calculation.
$\quad \blacksquare$
\vskip 0.4cm \noindent
This result determines $T_g$ for all symbols 
$g \in \mathbb{C}Q_q(\theta, \overline{\theta})  $ by linearity.
Also, this result exhibits $T_{\theta^i \overline{\theta}{}^j }$ as a
weighted shift operator with the degree of the shift being $i-j$.
Next to see when this operator is bounded or compact
we apply some basic functional analysis to obtain immediately:
\begin{corollary}
First, $T_{\theta^i \overline{\theta}{}^j }$ is a bounded operator if and
only if
$$
    || T_{\theta^i \overline{\theta}{}^j } ||_{op} =
     \sup \Big\{ \dfrac{w_{i+a}}{ (w_a \, w_{i+a-j} )^{1/2} }  \, \Big| \, a \ge 0 \Big\} < \infty,
$$
where $|| \cdot  ||_{op}$ denotes the operator norm.
Secondly, $T_{\theta^i \overline{\theta}{}^j }$ is a compact operator if and
only if
$$
         \lim_{a \to \infty} \dfrac{w_{i+a}}{ (w_a \, w_{i+a-j} )^{1/2} }  = 0.
$$
\end{corollary}
Knowing this, it is now easy to construct examples of Toeplitz operators which are not bounded
provided that we are free to choose the weights $w_k$.
Similarly, it is now straightforward to construct Toeplitz operators which are bounded, but not compact,
given the same freedom.
We also showed earlier that
 $T_1 = I_{Pre (\theta)}$, which is bounded but not compact.

\vskip 0.4cm \noindent
We next obtain a consequence which relates the adjoint of a Toeplitz operator with symbol $g$
to the Toeplitz operator with the conjugate symbol $g^*$.
\begin{theorem}
\label{theorem-Tg-and-Tg-star}
Let $g \in  \mathbb{C}Q_q(\theta, \overline{\theta})$ be arbitrary.
Then
\begin{equation}
\label{Tg-and-Tg-star}
\langle T_g f_1 , f_2 \rangle_w = \langle f_1, T_{g^*} f_2 \rangle_w
\end{equation}
for all $f_1, f_2 \in Pre(\theta)$.
\end{theorem}
\textbf{Proof:}
It suffices to prove this for $g = \theta^i \overline{\theta}{}^j$
where $i,j \ge 0$ and for
$f_1 = \phi_a$ and $f_2 = \phi_b$ where $a,b \ge 0$.
So we compute each side of (\ref{Tg-and-Tg-star}) for these choices.
For the left side we get
\begin{align}
\label{left-side}
\langle T_{\theta^i \overline{\theta}{}^j } \phi_a , \phi_b \rangle_w &=
\dfrac{w_{i+a}}{ (w_a \, w_{i+a-j} )^{1/2} } \langle \phi_{i+a-j}, \phi_b \rangle_w
\nonumber
\\
&=
\dfrac{w_{i+a}}{ (w_a \, w_{i+a-j} )^{1/2} } \delta_{i+a-j , b}.
\end{align}
Note that the Kronecker delta is enforcing the condition that
$i+a-j = b \ge 0$. 
Next for the right side we have
\begin{align}
\label{right-side}
\langle  \phi_a , T_{ (\theta^i \overline{\theta}{}^j)^* } \phi_b \rangle_w &=
\langle  \phi_a , T_{ \theta^j \overline{\theta}{}^i } \phi_b \rangle_w \nonumber
\\
&=
\dfrac{w_{j+b}}{ (w_b \, w_{j+b-i} )^{1/2} } \langle  \phi_a ,  \phi_{j+b-i}  \rangle_w \nonumber
\\
&=
\dfrac{w_{j+b}}{ (w_b \, w_{j+b-i} )^{1/2} } \delta_{a, j+b-i}.
\end{align}
This time the delta imposes the condition $j+b-i =a \ge 0$.
So in each case we have the combined conditions $a, b \ge 0$ and
$i+a = b+j$.
Using these conditions one sees that the expressions in (\ref{left-side}) and
(\ref{right-side}) are equal.
$\quad \blacksquare$

\vskip 0.4cm \noindent
\textbf{Remark:}
This result holds even when the inner product is degenerate.
However, even when the inner product is non-degenerate all it says
about the adjoint of $T_g$ is that $T_{g^*} \subset (T_g)^*$, that is,
the adjoint of $T_g$ is an extension of $T_{g^*}$.
Of course, such details are typical of densely defined operators.
We recall that the Toeplitz operators are densely defined operators, all of which
have the same dense domain, namely $Pre(\theta)$.
Also, this relation  $T_{g^*} \subset (T_g)^*$ shows a compatibility between 
  our definition of the conjugation in $ \mathbb{C}Q_q(\theta, \overline{\theta})$
and the adjoint of a Toeplitz operator.

\begin{corollary}
\label{symmetric-operator-corollary}
If $g \in  \mathbb{C}Q_q(\theta, \overline{\theta})$ is a self-adjoint element
(meaning $g^* =g)$, then $T_g$ is a symmetric operator.
\end{corollary}
\textbf{Proof:}
By Theorem \ref{theorem-Tg-and-Tg-star} and $g^* = g$ we have
\begin{equation*}
\langle T_g f_1 , f_2 \rangle_w = \langle f_1, T_{g} f_2 \rangle_w
\end{equation*}
for all $f_1, f_2 \in Pre(\theta) = Dom (T_g)$.
And this is exactly what it means for a densely defined operator to be symmetric.
(See \cite{reed-simon}.)
$\quad \blacksquare$

\vskip 0.4cm \noindent
\textbf{Remark:} If $g^* = g$, then it behooves us to study the self-adjoint extensions
of the symmetric operator $T_g$.
This remains an open problem.

\begin{corollary}
Every Toeplitz operator $T_g$ is closable and its closure satisfies
$\overline{T}_g = (T_g)^{**} \subset (T_{g^*})^*$.
\end{corollary}
\textbf{Proof:}
This follows rather directly from Theorem VIII.1b in \cite{reed-simon}.
We get from that reference that $T_g$ is closable if and only if
$Dom (T_g)^*$ is dense.
But $Dom (T_g)^* \supset Dom \, T_{g^*} = Pre (\theta) $ and $Pre (\theta) $ is dense.
The equality $\overline{T}_g = (T_g)^{**}$ follows from the cited theorem.
The inclusion $(T_g)^{**} \subset (T_{g^*})^*$ follows from Theorem \ref{theorem-Tg-and-Tg-star}.
$\quad \blacksquare$

\vskip 0.4cm \noindent
We now analyze various particular cases of (\ref{useful}).
First for $i = j = 0$ we have
$$
      (T_1 \phi_a)  (\theta) =  \dfrac{w_{a}}{ (w_a \, w_{a} )^{1/2} } \, \phi_{a}(\theta)
      = \phi_a (\theta)
$$
for all $a \ge 0$, so that $T_1 = I_{Pre(\theta)}$, the identity map, as already noted above.

For the case $i=j$ of (\ref{useful}) we obtain for all $a \ge 0$ that
$$
     (T_{\theta^i \overline{\theta}{}^i } \phi_a)    (\theta)
      =  \dfrac{w_{i+a}}{ (w_a \, w_{i+a-i} )^{1/2} } \, \phi_{i+a-i}(\theta) 
      =  \dfrac{w_{i+a}}{ (w_a \, w_{a} )^{1/2} } \, \phi_{a}(\theta) 
      =  \dfrac{w_{i+a}}{ w_{a} } \, \phi_{a}(\theta) .
$$
Hence the basis $\phi_{a}(\theta)$ diagonalizes simultaneously the family of symmetric
operators $T_{\theta^i \overline{\theta}{}^i }$ for $i \ge 0$.
By Corollary \ref{symmetric-operator-corollary} we see that 
$T_{\theta^i \overline{\theta}{}^i }$  is symmetric.

Next we consider (\ref{useful}) for the case $j=0$ and get
$$
         (T_{\theta^i  } \phi_a)    (\theta) = \dfrac{w_{i+a}}{ (w_a \, w_{i+a} )^{1/2} } \, \phi_{i+a}(\theta) 
         = \dfrac{w_{i+a}^{1/2}  }{ w_a^{1/2} } \, \phi_{i+a}(\theta)
$$
or, equivalently,
$
        T_{\theta^i } : \theta^a \mapsto \theta^{i+a}
$
which itself can be written as $T_{\theta^i } = M_{\theta^i } $.
Of course, this also follows from the definition $T_{\theta^i } = P_K M_{\theta^i }  = M_{\theta^i } $,
since $M_{\theta^i } $ leaves $Pre (\theta)$ invariant and $P_K$ acts as the identity on $Pre (\theta)$.
A subcase here is $T_{\theta } = M_{\theta } $, which merits the name \textit{creation operator}
since it increases by $1$ the degree of the elements in $Pre (\theta)$, which are exactly the polynomials
in $\theta$.
Moreover, $T_{\theta^i } = (T_{\theta})^i$ also is immediate. 
(Recall that $T_{\theta}$ leaves $Pre(\theta)$ invariant, and so $(T_{\theta})^i$ is defined.)
So, if $T_{\theta}$ is bounded (resp., compact), then $T_{\theta^i } $ is bounded
(resp., compact) for all $i \ge 1$.
In the Hilbert space introduced by Bargmann in \cite{BA}, one has $w_a = a!$ and $\theta = z$, so that 
$T_{\theta^i } = T_{z^i}$ is not bounded for $i \ge 1$ in that space.
One expects that with $w_a$ being some reasonable deformation of the factorial function
the corresponding operators $T_{\theta^i } $ would also not be bounded.
However, the boundedness of these operators depends completely on the choice of weights
$w_a$, nothing else. 
So for some choices (such as, for example, $w_a$ constant) these operators will be bounded.

Yet another interesting special case of  (\ref{useful}) is when $i=0$.
Then we have
$$
  (T_{ \overline{\theta}{}^j } \phi_a)    (\theta) = \dfrac{w_{a}}{ (w_a \, w_{a-j} )^{1/2} } \, \phi_{a-j}(\theta)
  = \left( \dfrac{w_{a}}{ w_{a-j} } \right)^{1/2}  \, \phi_{a-j}(\theta)
$$
or, in terms of the unnormalized monomials,
$$
        T_{ \overline{\theta}{}^j } : \theta^a \mapsto \dfrac{w_a}{w_{a-j}} \, \theta^{a-j}
$$
for all $ a \ge 0$.
In particular, for $j=1$ we see that 
$$
        T_{ \overline{\theta} } : \theta^a \mapsto \dfrac{w_a}{w_{a-1}} \, \theta^{a-1}
$$
deserves to be called an \textit{annihilation operator}, since it lowers the degree of
any non-constant polynomial by $1$ and sends constants to zero.
A simple argument shows that $T_{ \overline{\theta}{}^j } =  ( T_{ \overline{\theta} } )^j $.
And similar to the above situation, we see that if $T_{ \overline{\theta} } $ is bounded (resp., compact),
then $T_{ \overline{\theta}{}^j }$ is bounded (resp., compact) for all $j \ge 1$.
Again, the space in \cite{BA} is an important example for which the operators 
$T_{ \overline{\theta}{}^j } $ are not bounded. 
And again, the boundedness of these operators depends solely on the weights.

Using Theorem \ref{nice-theorem} in the first equality and two properties established above in the second
equality, we see that
$$
T_{\theta^i \overline{\theta}{}^j } = T_{\overline{\theta}{}^j} T_{\theta^i } = 
 (T_{\overline{\theta}})^j (T_{\theta})^i.
$$
The last expression here is in anti-Wick order, which by definition means a product of creation and
annihilation operators such that
all of the creation operators are to the right of all of the annihilation operators.
By linearity every Toeplitz operator 
$T_g$ will then be a sum of terms, each of which is in anti-Wick order.
Because of this property one says that the Toeplitz quantization is an
\textit{anti-Wick quantization}.

There is another way of viewing the annihilation operator $T_{ \overline{\theta} } $.
We note that in the case when $w_a = a!$ as in \cite{BA}, we have that 
$$
        T_{ \overline{\theta} } : \theta^a \mapsto \dfrac{w_a}{w_{a-1}} \, \theta^{a-1} = 
        \dfrac{a!}{ (a-1)! } \, \theta^{a-1} = a \,  \theta^{a-1} ,
$$
which is the derivative operator from elementary calculus.
So we can think of $T_{ \overline{\theta} }$ in this more general context as a
deformation of the classical derivative.
We call it the \textit{$w$-deformed derivative} and denote it by $\partial_w$.
If we define the $w$-deformed integers to be $[n]_w := w_n / w_{n-1}$ for every integer $n \ge 1$
and $[0]_w := 0$, then we have 
$$
        \partial_w = T_{ \overline{\theta} } : \theta^a \mapsto [a]_w \, \theta^{a-1}.
$$
The upshot of this paragraph is merely a change to a notation that is
more compatible with notations used elsewhere in the literature, nothing else really.

Notice again that $T_{\overline{\theta}{}^j} T_{\theta^i } = T_{\theta^i \overline{\theta}{}^j }$ follows
from Theorem \ref{nice-theorem}.
We now calculate $T_{\theta^i } T_{\overline{\theta}{}^j} $ using the individual formulas derived
above for $T_{\theta^i }$ and $ T_{\overline{\theta}{}^j} $.
So,
\begin{equation*}
\phi_a \stackrel{T_{\overline{\theta}{}^j} }{\longrightarrow} \left( \dfrac{w_a}{w_{a-j}} \right)^{1/2} \phi_{a-j}
 \stackrel{ T_{\theta^i } }{\longrightarrow} 
 \left( \dfrac{w_a}{w_{a-j}} \right)^{1/2}   \left( \dfrac{ w_{i+a-j} }{ w_{a-j} } \right)^{1/2} \phi_{a-j+i} 
\end{equation*}
which gives
$$
      T_{\theta^i } T_{\overline{\theta}{}^j} \phi_a = \dfrac{{ ( w_a w_{i+a-j} )^{1/2} }}{w_{a-j}} \phi_{a-j+i}.
$$
This is different from the formula (\ref{useful}) derived above for $ T_{\theta^i \overline{\theta}{}^j }$.
In particular, for the case $i=j=1$ which we left unfinished earlier we have
$$
      T_{\theta } T_{\overline{\theta}} \phi_a = \dfrac{{ w_{a}  }}{w_{a-1}} \phi_{a} = [a]_w  \phi_{a}.
$$
For the sake of completeness we note that
the operator $N_\theta := T_{\theta } T_{\overline{\theta}}$ is called the 
\textit{$w$-deformed number operator}.
On the other hand from equation (\ref{useful}) we have that
$$
    T_{\overline{\theta}} T_{\theta } \phi_a = T_{\theta \overline{\theta} } \phi_a = \dfrac{w_{a+1}}{w_a} \phi_a
    = [a+1]_w  \phi_{a}.
$$

\section{Canonical Commutation Relations}

This final section is a continuation of the two calculations just made at the end of the last section.
First, we define the $q$-commutator of any two elements $a$ and $b$ in any
(associative, say) algebra over $\mathbb{C}$ by
$$
      [ a , b ]_q := a b -  q b a,
$$
where $q \in \mathbb{C} \setminus \{ 0 \}$.
This is the commutator which is appropriate for the study
of $q$-deformations.

The Toeplitz quantization starts with the `classical' space  $\mathbb{C}Q_q ( \theta, \overline{\theta} )$
of symbols and produces operators  acting in the `quantum' Segal-Bargmann space $\mathcal{B}(\theta)$.
The point here is that before the Toeplitz quantization we have the \textit{homogeneous}
$q$-commutation relation in $\mathbb{C}Q_q ( \theta, \overline{\theta} )$, namely
\begin{equation}
\label{theta-ccr}
         [ \theta , \overline{\theta}  ]_q =  \theta \overline{\theta} - q \overline{\theta} \theta = 0.
\end{equation}

Speaking roughly without going into the rigorous details,
in quantum theory we have creation operators and annihilations operators 
which come in pairs, say $A$ for an annihilation operator and $A^*$ for its corresponding
creation operator.
Then a typical commutation relation is something more or less like
$$
           [ A, A^*] = I, \quad \mathrm{the~identity.}
$$
This is called a \textit{canonical commutation relation}.
So in general in a quantum theory we expect \textit{inhomogeneous}
canonical commutation relations. 

Now the Toeplitz quantization of the $q$-commutator $ [ \theta , \overline{\theta}  ]_q $ is
$$
     [ T_\theta, T_{\overline{\theta} } ]_q =   T_\theta T_{\overline{\theta} }- q T_{\overline{\theta}} T_\theta.
$$
But recall that $T_\theta$ is the creation operator and that $T_{\overline{\theta} }$
is the annihilation operator; so this $q$-commutator has the form $[A^*, A]_q$.
And this is not the form of a canonical commutation relation.
However, since it is homogeneous and $q \ne 0$ we can trivially rewrite 
(\ref{theta-ccr}) as
\begin{equation}
\label{theta-ccr-rewrite}
         [ \overline{\theta}, \theta  ]_{ q^{-1} } = \overline{\theta} \theta - q^{-1} \theta \overline{\theta} = 0.
\end{equation}
In fact we have an identification
$\mathbb{C}Q_q ( \theta, \overline{\theta} ) \cong \mathbb{C}Q_{q^{-1}} ( \overline{\theta}, \theta )$.
What this means is that at the classical level we can not distinguish the $q$-deformed theory
associated to the holomorphic (resp., anti-holomorphic) variable $\theta$ (resp., $\overline{\theta}$)
from the $q^{-1}$-deformed theory
associated to the holomorphic (resp., anti-holomorphic) variable $\overline{\theta}$ (resp., $\theta$).
(The previous sentence does not contain a typographical error. It makes perfect sense to consider
 $\overline{\theta}$ as a holomorphic variable whose associated anti-holomorphic variable
 is $\theta$.)
 Another way of saying this is that as far as our theory is concerned only with the classical level
 we have no way to distinguish between $q$-deformations and $q^{-1}$-deformations nor
 between holomorphic and anti-holomorphic variables.
 
 However, the quantizations of $\theta$ and $\overline{\theta}$ are distinguishable. 
 In this sense Toeplitz quantization breaks a symmetry. 
 And the choice of quantization determines exactly how the symmetry is broken.
 For example, if we define a Toeplitz quantization as in this paper, but using instead the anti-Segal-Bargmann
 space $\mathcal{B}(\overline{\theta})$ as the Hilbert space 
 in which the quantized operators act, then $\theta$ quantizes to 
 the annihilation operator while $\overline{\theta}$ quantizes to 
 the creation operator, just the reverse of what we have obtained with the present Toeplitz quantization in
 the Segal-Bargmann space $\mathcal{B}(\theta)$.
 These comments indicate that naming a particular order in 
 $\mathbb{C}Q_q ( \theta, \overline{\theta} ) $
 the anti-Wick ordering (that is, all creation operators to the right of all annihilation operators) is not
 justifiable in terms of mathematical structures of $\mathbb{C}Q_q ( \theta, \overline{\theta} ) $ alone.
 We have simply decided to follow the nomenclature used in \cite{csq} as indicated earlier.
  
 Now the Toeplitz quantization of the $q^{-1}$-commutator $ [ \overline{\theta}, \theta  ]_{ q^{-1} }$ is
 $$
  [ T_{\overline{\theta}}, T_\theta  ]_{ q^{-1} } = T_{\overline{\theta}} T_\theta - q^{-1} T_\theta T_{\overline{\theta}}.
 $$
And this has the virtue of being of the form $[A, A^*]$.
So we require this canonical commutation relation to hold:
\begin{equation}
 [ T_{\overline{\theta}}, T_\theta  ]_{ q^{-1} } = T_{\overline{\theta}} T_\theta - q^{-1} T_\theta T_{\overline{\theta}}
  = I_{Pre(\theta)}, \quad \mathrm{the~identity~on~} Pre(\theta).
\end{equation}
This gives us the recursion relation
$$
   [a +1 ]_w - q^{-1}  [a]_w = 1
$$
for all $a \ge 0$.
But we already have $[0]_w =0$.
So the sequence $[a]_w$ is uniquely determined by $q$
(or by $ q^{-1}$ if one wishes to consider this as the primary parameter).
It is rather straightforward to find an explicit formula for  $[a]_w$.
The next definition is standard, though not universal.
See \cite{csq} for a different, more symmetric definition.

\begin{definition}
Let $r \in \mathbb{C}$. For each integer $n \ge 0$ we define
$$
      [ n ]_r := 1 + r + r^2 + \cdots + r^{n-1} \quad \mathrm{if} \,\,\, n \ge 1
$$
and $[0]_r := 0$.
This is called the {\rm $r$-deformation} of $n$.
\end{definition}
For example, $[1]_r = 1$ and $[2]_r = 1 + r $.
Taking $r=1$ gives $[ n ]_r =n$ for every integer $n \ge 0$.
This justifies saying that these are deformations of the integers and
that $r$ in the deformation parameter.
If $r \ne 1$, then we have the alternative expression $[ n ]_r = \frac{1 - r^n}{1 - r}$,
 which often appears in the literature.
\begin{prop}
The unique solution of the recursion relation 
$$ 
[a +1 ]_w - q^{-1}  [a]_w = 1
$$
for all integers $a \ge 0$
with $[0]_w =0$ is
$[ a ]_w = [ a ]_{ q^{-1} }$.
\end{prop}
\textbf{Proof:} 
The recursion relation for $[n]_r$ is $[n+1]_r - r [n]_r =1$, as the reader
can easily check.
Taking $r=q^{-1}$ shows that the sequences $  [ a ]_w$ and  $[ a ]_{ q^{-1} }$
satisfy the same recursion relation.
But they both start out with $ [ 0 ]_w = 0 = [ 0 ]_{ q^{-1} }$, which ends the proof.
$\quad \blacksquare$

Now it is a matter of going from the deformed integers $   [ a ]_w = [ a ]_{ q^{-1} }$ to the
weights $w_k$.
Now for every integer $a \ge 1$ we have
\begin{equation}
\label{a-w}
         [ a ]_{ q^{-1} } =  [ a ]_w = \dfrac{w_a}{w_{a-1}}
\end{equation}
by definition of $[ a ]_w$.
It turns out that  $[ 0 ]_w = 0$ carries no information about the weights.
Then (\ref{a-w}) gives a sequence of identities 
\begin{equation*}
w_1 = [1]_{ q^{-1} } w_0,
\qquad
w_2 = [2]_{ q^{-1} } w_1,
\qquad
w_3 = [3]_{ q^{-1} } w_2,
\end{equation*}
and so on.
The solution for $k \ge 1$ is clearly
$$
       w_k =  [k]!_{ q^{-1} } w_0,
$$
where the $q^{-1}$-deformed factorial is defined by 
$$
 [k]!_{ q^{-1} } := [k]_{ q^{-1} } [k-1]_{ q^{-1} } \cdots [2]_{ q^{-1} } [1]_{ q^{-1} }
$$
and where $w_0 > 0$ is arbitrary.
In this way we have defined a unique sequence (up to a multiplicative positive constant) 
of weights $w_k = w_k(q)$, which are functions of the one parameter $q$ such that
$$
  [ T_{\overline{\theta}}, T_\theta  ]_{ q^{-1} }
   = T_{\overline{\theta}} T_\theta - q^{-1} T_\theta T_{\overline{\theta}}  =I_{Pre(\theta)}.
$$
In particular, $[ T_{\overline{\theta}}, T_\theta  ]_{ q^{-1} }$ is bounded.
By putting the deformation parameter $q$ equal to $1$ and normalizing 
$1 \in \mathbb{C}Q_q ( \theta, \overline{\theta} )$ by putting
$w_0 =1$, we recover the weights $w_k = k!$ of the Hilbert space
$\mathcal{H}$ defined in (\ref{theHilbertspace}).
Recall that the Segal-Bargmann space based on the phase space $\mathbb{C}$
in \cite{BA} is the closed subspace of $\mathcal{H}$ consisting of
the holomorphic functions in $\mathcal{H}$.

If we wish to have some other operator instead of the identity
on the `right side' of the canonical
commutation relation, the same method applies to give the corresponding weights.

\section{Concluding Remarks}

Since the Toeplitz operators introduced here are only densely defined, one has the standard
problems in the analysis of such operators. 
For example, we know they are closable, but can we identify exactly what the closure is?
And if a Toeplitz operator is symmetric, 
then we would like to know what its self-adjoint extensions are.
In particular, we would like to know exactly what are the conditions for
a Toeplitz operator to be essentially self-adjoint.

We have given necessary and sufficient conditions for the Toeplitz
$T_{\theta^i \overline{\theta}{}^j}$ to be bounded or compact. 
But the full story remains to be told 
for $T_g$ where $g$ is an arbitrary symbol, though our results allow us to form
sufficient conditions for boundedness and compactness by expanding $T_g$ as
a linear combination of $T_{\theta^i \overline{\theta}{}^j}$'s.
We expect such conditions to be far from necessary.

Another possibility for further research is to define coherent states in this context,
much as was done in \cite{csq} in a similar finite dimensional case.
This would allow the introduction of a coherent state transform and a 
coherent state quantization.
(Also see \cite{gazeau}.)
This would relate the material in this paper with yet another aspect of mathematical physics.
Also it might be of interest to study in more detail the classical space
 $\mathbb{C}Q_q ( \theta, \overline{\theta} ) $ from a physics point of view
 as a sort of non-commutative phase space.
 
 Given the positive result in the finite dimensional case presented in \cite{part1}
 it seems reasonable to conjecture that $\mathbb{C}Q_q ( \theta, \overline{\theta} ) $
 also has its own reproducing kernel, at least in the case when its inner product
 is non-degenerate.
 We also leave this as a problem for another day.

\vskip 0.4cm \noindent
\textbf{Acknowledgments}:
I am most grateful to Jean-Pierre Gazeau  for drawing my attention to the paper \cite{csq}
and for various illuminating comments about it.
He ignited the spark which started the flame which burns to this day.
Merci beaucoup, Jean-Pierre!
I also am glad to thank Gabriel Kant\'un-Montiel for helping me clarify several points.
Muchas gracias, Gabriel!

\end{document}